\documentclass{aastex631}

\usepackage{verbatim}
\usepackage{amsmath,bm}
\usepackage{color}
\usepackage{courier}
\usepackage{threeparttable}
\usepackage{ragged2e}
\usepackage{tablefootnote}
\usepackage{hyperref}
\usepackage{footmisc}

\newcommand{\natas}{{\it Nature Astronomy}}

\newcommand{\ha}{\hbox{H$\alpha$}}
\newcommand{\hb}{\hbox{H$\beta$}}

\newcommand{\oiii}{\hbox{[O\,{\sc iii}]}}
\newcommand{\nii}{\hbox{[N\,{\sc ii}]}}
\newcommand{\sii}{\hbox{[S\,{\sc ii}]}}

\shorttitle{A dual active black hole candidate}
\shortauthors{Cao et al.}
\graphicspath{{./}{figures/}}

\begin{document}

\title{A Dual Active Black Hole Candidate with Mass Ratio $\sim$7:1 in a Disk Galaxy}

\author{Xiao Cao}
\affiliation{School of Astronomy and Space Science, Nanjing University, Nanjing 210023, People's Republic of China}
\affiliation{Key Laboratory of Modern Astronomy and Astrophysics (Nanjing University), Ministry of Education, Nanjing 210023, People's Republic of China}

\author{Yan-Mei Chen}
\affiliation{School of Astronomy and Space Science, Nanjing University, Nanjing 210023, People's Republic of China}
\affiliation{Key Laboratory of Modern Astronomy and Astrophysics (Nanjing University), Ministry of Education, Nanjing 210023, People's Republic of China}

\author{Yong Shi}
\affiliation{School of Astronomy and Space Science, Nanjing University, Nanjing 210023, People's Republic of China}
\affiliation{Key Laboratory of Modern Astronomy and Astrophysics (Nanjing University), Ministry of Education, Nanjing 210023, People's Republic of China}

\author{Junfeng Wang}
\affiliation{Department of Astronomy, Xiamen University, Xiamen 361005, People's Republic of China}

\author{Zhi-Jie Zhou}
\affiliation{School of Astronomy and Space Science, Nanjing University, Nanjing 210023, People's Republic of China}
\affiliation{Key Laboratory of Modern Astronomy and Astrophysics (Nanjing University), Ministry of Education, Nanjing 210023, People's Republic of China}

\author{Min Bao}
\affiliation{School of Astronomy and Space Science, Nanjing University, Nanjing 210023, People's Republic of China}
\affiliation{Key Laboratory of Modern Astronomy and Astrophysics (Nanjing University), Ministry of Education, Nanjing 210023, People's Republic of China}

\author{Qiusheng Gu}
\affiliation{School of Astronomy and Space Science, Nanjing University, Nanjing 210023, People's Republic of China}
\affiliation{Key Laboratory of Modern Astronomy and Astrophysics (Nanjing University), Ministry of Education, Nanjing 210023, People's Republic of China}

\author{Alexei Moiseev}
\affiliation{Special Astrophysical Observatory, Russian Academy of Sciences, Nizhny Arkhyz, Russia}
\affiliation{Sternberg Astronomical Institute, Lomonosov Moscow State University, Moscow, Russia}

\author{Luis C. Ho}
\affiliation{Kavli Institute for Astronomy and Astrophysics, Peking University, Beijing 100871, People's Republic of China}
\affiliation{Department of Astronomy, School of Physics, Peking University, Beijing 100871, People's Republic of China}

\author{Lan Wang}
\affiliation{National Astronomical Observatory, Chinese Academy of Sciences, Beijing 100101, People's Republic of China}
\affiliation{School of Astronomy and Space Science, University of Chinese Academy of Sciences, Beijing 100049, People's Republic of China}

\author{Guangquan Zeng}
\affiliation{National Astronomical Observatory, Chinese Academy of Sciences, Beijing 100101, People's Republic of China}
\affiliation{School of Astronomy and Space Science, University of Chinese Academy of Sciences, Beijing 100049, People's Republic of China}

\correspondingauthor{Yan-Mei Chen}
\email{chenym@nju.edu.cn}

\begin{abstract}
    Dual active galactic nuclei (AGNs) with comparable masses are commonly witnessed among the major merged galaxies with interaction remnants. Considering almost every massive galaxy is associated with multiple dwarf satellites around it, minor mergers involving galaxies with disproportional stellar masses should be much more common than major mergers, which would naturally lead to black hole (BH) pairs with significantly different masses. However, dual AGNs generated by minor mergers involving one or two dwarf galaxies are exceptionally rare and understudied. Moreover, good estimates of the masses of both BHs are not yet available to test this idea. Here we report the evidence of a dual AGN candidate with mass ratio $\sim$7:1 located in an undisturbed disk galaxy. We identify the central BH with mass of $9.4 \times 10^6M_\odot$ from its radio emission as well as AGN-driven galactic-scale biconical outflows. The off-centered BH generates obvious broad and narrow emission-line regions, which gives us a robust estimation of a $1.3 \times 10^6M_\odot$ BH mass. We explore alternative scenarios for explaining the observational features of this system, including the complex gas kinematics triggered by central AGN activity and dust attenuation of the broad-line region of the central BH, finding that they failed to fully account for the kinematics of both the redshifted off-centered broad and narrow emission-line components.
\end{abstract}


\section{Introduction} \label{sec:intro}

Galaxy mergers are ubiquitous and violent phenomena in the Universe, during which the interactions and/or collisions between two/multiple galaxies transform galaxy morphology, reshape the stellar and gaseous kinematics, and affect galaxy assembly histories. Since almost every massive galaxy is believed to harbor a massive black hole (BH), galaxy mergers are logical precursors of dual massive BH systems that could be detected if gas fuel from the interstellar medium (ISM) triggers the activities of both BHs \citep{2008ApJS..175..356H,2015ApJ...806...22D,2018ApJ...856....6D}. Over the past 20 yr, dual active galactic nuclei (AGNs) have been of great interest because they provide a direct probe of the coevolution between massive galaxies and their central BHs, and represent unique laboratories for exploring physical processes driven by mergers, including star formation, evolution of chemical composition, AGN activity, and gravitational waves.

The connection between the incidence of dual AGNs and major mergers (mass ratio $<$3:1), involving two large galaxies with comparable masses, has been studied extensively \citep{2003ApJ...582L..15K,2008MNRAS.386..105B,2011ApJ...733..103F,2012ApJ...753...42C,2012ApJ...746L..22K,2013ApJ...762..110L,2022NatAs...6.1185M}. In the early stages, systematic searches for kiloparsec-scale dual AGNs have focused on AGNs with double-peaked narrow emission lines. Based on the single fiber spectral sample, \cite{2009ApJ...705L..76W}, \cite{2010ApJ...708..427L} and \cite{2010ApJ...716..866S} selected AGNs with double-peaked narrow emission lines as dual AGN candidates from Data Release 7 of the Sloan Digital Sky Survey \citep[SDSS;][]{2009ApJS..182..543A}, which in total contained 340 galaxies. However, follow-up multiwavelength observations for subsamples of these double-peaked objects found that double-peak selection is not an efficient method for searching for dual AGNs \citep{2010ApJ...716..131R,2011ApJ...739...44R,2011ApJ...740L..44F,2011ApJ...735...48S,2011AJ....141..174T,2015ApJ...811...14M}. From 18 double-peaked narrow-line AGNs having detections by Faint Images of the Radio Sky at Twenty-Centimeters \citep[FIRST;][]{1995ApJ...450..559B} survey and exhibiting two compact AGN emission components with angular separations larger than 0.2$''$ in optical long-slit spectra, \cite{2015ApJ...813..103M} found $\sim$17\% (3/18) galaxies have two spatially separated radio cores in images and suggested them as dual AGNs. Combining integral field spectroscopy (IFS) and high-resolution imaging, \cite{2012ApJ...745...67F} investigated 106 double-peaked AGNs, finding only $\sim$2\% (2/106) of these galaxies are dual AGNs. The dual AGN with the smallest known separation so far is $\rm UGC\ 4211$. This system was originally infrared selected (in the late 1980s) and identified as an optical AGN in follow-up spectroscopy. Recently, with the colocated observations of high-spatial-resolution images from Hubble Space Telescope (HST) F814W, Keck $J$ and $K'$ bands, ALMA continuum at $\sim$230 GHz, as well as the optical IFS from MUSE, \cite{2023ApJ...942L..24K} identified a separation of 230 pc between the two AGNs in $\rm UGC\ 4211$ and estimated the masses of the two BHs as ${\rm log}\ (M_{\rm BH}/M_\odot)\sim 8.1$ and ${\rm log}\ (M_{\rm BH}/M_\odot)\sim 8.3$. Most of these systems mentioned above are powered by BHs with similar masses, indicating that they are remnants of major mergers and offering valuable laboratories to study the physics of the growth and accretion history of SMBHs.

Given that almost every massive galaxy is surrounded by multiple dwarf satellites, minor mergers involving galaxies with disproportional stellar masses (mass ratio $>$3:1) are expected to be much more frequent than major mergers. This naturally leads to the formation of BH pairs with significantly different masses. However, dual AGNs in minor mergers involving one or more dwarf galaxies are exceptionally rare and remain largely understudied \citep{2003ApJ...597..823C,2012ApJ...746L..22K,2015ApJ...806..219C,2017ApJ...836..183S,2018ApJ...862...29L}. Theoretical studies predict that both major and minor mergers are expected to be important for dual AGN growth \citep{2015MNRAS.447.2123C,2023MNRAS.522.1895C}. Different hydrodynamical simulations found that (1) the activity duration of dual BHs decreases with increasing mass ratio \citep{2017MNRAS.469.4437C}, and (2) the ram pressure stripping at the early stage and the tidal force at the late stage of the merger process suppress effective gas accretion onto the BH with lower mass in the dual system \citep{2011ApJ...729...85C,2012ApJ...748L...7V,2015MNRAS.447.2123C}. These simulations converge to a conclusion that the activities of two BHs in most minor mergers are not simultaneous as those shown in major mergers, explaining the scarcity of dual AGNs with disproportional masses.

In this work, we investigate an edge-on disk galaxy SDSS J144515.46+492605.4 (hereafter SDSS J1445+4926 for simplicity) based on the Mapping Nearby Galaxies at Apache Point Observatory (MaNGA) Survey \citep{2015ApJ...798....7B}. This galaxy is intriguing in its galactic-scale biconical outflow, and an abnormal redshifted gas kinematics located $\rm\sim 1\ kpc$ away from the galaxy photometric center. In Section \ref{sec:data}, we present the properties of SDSS J1445+4926 as well as the data analysis method. In Section \ref{sec:origin}, we discuss possible origins of the abnormal redshifted kinematics, including an off-centered BH, the central AGN activity-triggered complex gas kinematics, and the effect of dust extinction. We prefer that the abnormal redshifted kinematics is induced by an off-centered BH originating from a minor merger. Follow-up observations with high-resolution multi-waveband imaging and spectra are required to confirm this picture. Section \ref{sec:discuss} is the conclusion of this work.

\section{Data analysis} \label{sec:data} 
\subsection{MaNGA Data} \label{sec:manga}

MaNGA \citep{2015ApJ...798....7B,2015AJ....149...77D,2017AJ....154...86W} is one of the projects in the fourth generation SDSS survey \citep[SDSS-IV;][]{2017AJ....154...28B}. It provides 3600$-$10,300\AA\ spatially resolved spectra with a median resolution of $R$$\sim$2000, using the Baryon Oscillation Spectroscopy Survey (BOSS) spectrographs \citep{2013AJ....146...32S} on the 2.5 m Sloan Telescope \citep{2006AJ....131.2332G}. MaNGA selects Primary and Secondary samples defined by two radial coverage goals. The Primary sample extends to a 1.5 effective radius ($R_e$, Petrosian 50\% light radius), while the Secondary sample extends to 2.5$R_e$ \citep{2016AJ....152..197Y}. SDSS J1445+4926 belongs to the Secondary sample with $R_e \sim 7.36''$ in $r$-band. The 1$''$ angular size corresponds to 0.593 kpc at the redshift $z\sim0.0296$ of this galaxy. The median guider seeing of the observation for SDSS J1445+4926 is $\sim$1.35$''$ obtained from the MaNGA data reduction pipeline \citep[DRP;][]{2016AJ....152...83L} catalog\footnote{\url{https://data.sdss.org/sas/dr17/manga/spectro/redux/v3_1_1/}}. The spatially resolved properties from integral field unit (IFU) observation within a 27$''$ diameter ``hexabundle," e.g., gas/stellar velocity fields, flux, and equivalent width of emission lines, are measured by the MaNGA data analysis pipeline \citep[DAP;][]{2019AJ....158..231W}. The MaNGA DAP used single Gaussian component to model each emission line and all the emission lines are tied to have the same central velocity.

\begin{figure}[ht!]
    \centerline{ \includegraphics[width=0.98\textwidth]{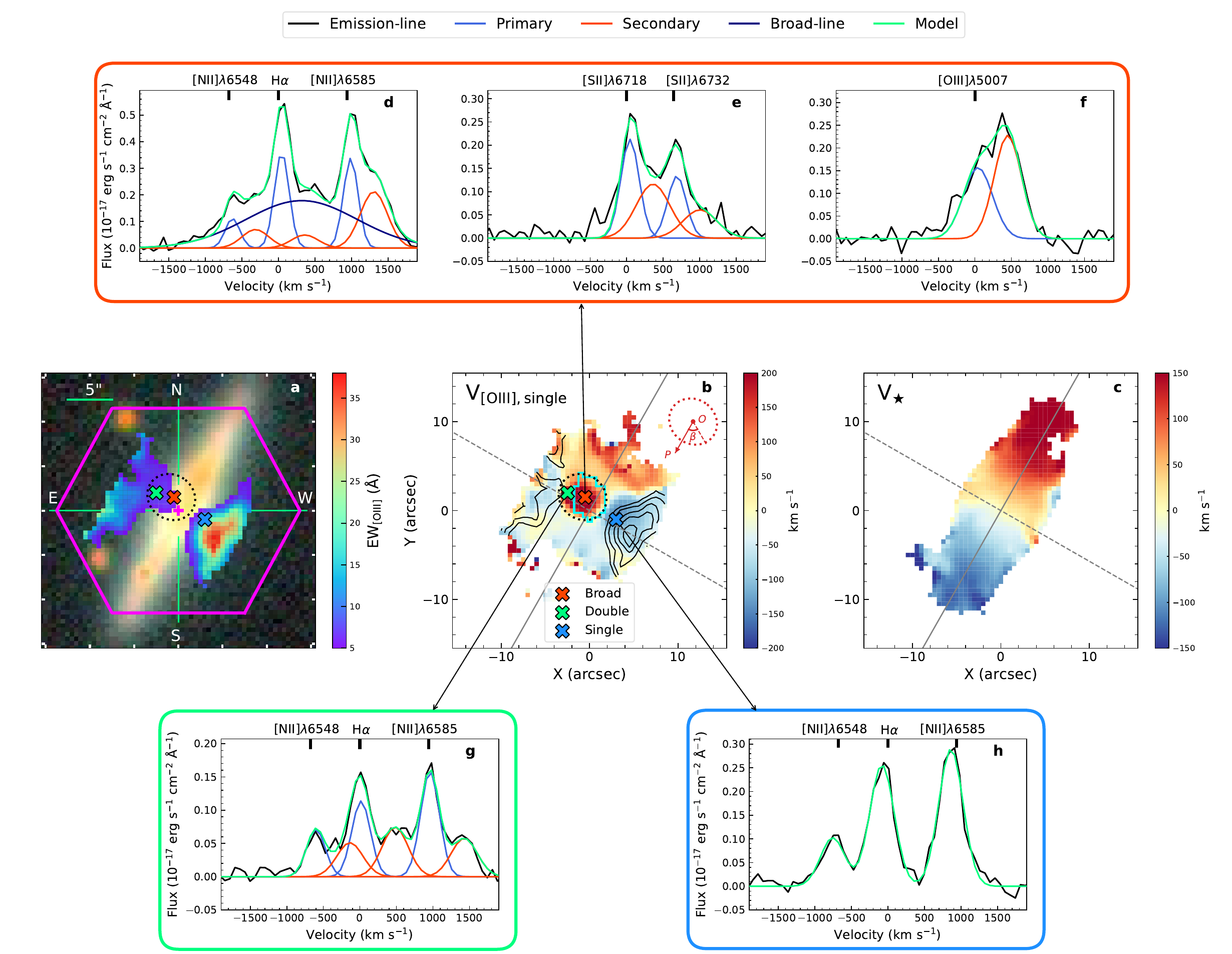} }
    \vspace{4mm}
    \caption{{\bf Morphology, kinematics and emission-line measurements of SDSS J1445+4926.}
    {\bf (a)} The $g$-, $r$-, and $z$-band composite image from DESI overlaid with the nonparametric EW$_{\rm [OIII]}$ map, the purple hexagon represents the MaNGA bundle size and the purple cross marks the galaxy photometric center.
    {\bf (b)} Gas velocity field. Black contours represent 20\%, 30\%, 50\%, 70\% and 90\% levels of EW$_{\rm [OIII]}$. The black dotted ellipse in panels (a),(b) marks an off-centered region with abnormal velocities (``Region A" in the text). The cyan polygon marks a region requiring \ha\ broad components for emission-line modeling.
    {\bf (c)} Stellar velocity field. The gray solid (dashed) line is the photometric major (minor) axis.
    {\bf (d) to (f)} \ha+\nii, \sii\ doublet and \oiii$\lambda$5007 emission-line fitting results for a spaxel marked as a red cross in panel (b). Black is the observed emission-line spectrum, 
    blue and red show the two narrow Gaussian components, and the dark-blue Gaussian in panel (d) is applied to model the broad \ha\ component. Green is the best-fitting model$-$it is a combination of all the Gaussian components.
    {\bf (g)} \ha+\nii\ fitting for the spaxel marked by the green cross in panel (b).
    {\bf (h)} \ha+\nii\ wavelength range fitted by a single Gaussian model for the spaxel marked as the blue cross in panel (b).}
    \label{fig:morph}
    \vspace{2mm}
\end{figure}

Based on the distribution of \oiii$\lambda$5007 equivalent width (EW$_{\rm [OIII]}$), \cite{2024MNRAS.531.2462Z} selected 142 outflow candidates with enhanced EW$_{\rm [OIII]}$ along the photometric minor axis from 10,010 unique galaxies observed by MaNGA. Figure \ref{fig:morph} shows SDSS J1445+4926 as an example with a clear biconical ionized structure, which is an edge-on disk galaxy with $\rm log\ sSFR/yr^{-1}=-10.15$, where ${\rm sSFR\equiv SFR}/M_\star$. The global star formation rate (SFR) is $\sim$0.56 $M_\odot\rm\ yr^{-1}$ given by \cite{2024arXiv240901279J} and the global stellar mass ($M_\star$) is $\sim$$10^{9.9}\ M_\odot$ taken from NASA-Sloan Atlas \citep[NSA;][]{2011AJ....142...31B}. Figure \ref{fig:morph}(a) presents the $g$-, $r$-, and $z$-band composite image from the Dark Energy Spectroscopic Instrument (DESI) Legacy Survey \citep{2019AJ....157..168D}. The overlaid color map shows the nonparametric EW$_{\rm [OIII]}$. Figure \ref{fig:morph}(b) shows  the \oiii$\lambda$5007 velocity field, where EW$_{\rm [OIII]}$ is overlaid as black contours. SDSS J1445+4926 is notable in the outflow candidates for two reasons: (1) it is clear the EW$_{\rm [OIII]}$ enhanced biconical regions have gas kinematics largely deviating from a regular rotating disk, and it is outflow dominated; (2) there is an abnormal redshifted gas component marked by the black dotted ellipse in \oiii$\lambda$5007 velocity field$-$we refer to this region as ``Region A" in the following, and the coordinate of its geometric center is listed in Table \ref{tab:1}. We also present Region A as a black dotted ellipse in the DESI image. The stellar velocity field is shown in Figure \ref{fig:morph}(c). The gray solid (dashed) lines in gaseous and stellar velocity fields mark the photometric major (minor) axis taken from the NSA catalog. 

In order to explore the origin of the redshifted gaseous component within Region A, we check the spectrum of each spaxel, finding that the emission-line spectra cannot be modeled by a single Gaussian component for spaxels within Region A. For the forbidden lines such as \oiii$\lambda$5007, \nii$\lambda\lambda$6548,6585 and \sii$\lambda\lambda$6718,6732, double Gaussian components are required. In addition to the double narrow components, an extra broad component is necessary for \ha\ emission. We show emission-line fitting examples for three spaxels in Figures \ref{fig:morph}(d)-(h). The details of the fitting process are presented in Section \ref{sec:method}. Figures \ref{fig:morph}(d)-(f) show an example of \ha+\nii, \sii\ and \oiii\ spectra for the geometric center of Region A marked by the red cross in Figures \ref{fig:morph}(a) and (b). Black is the observed emission-line spectrum, blue and red show the two narrow components, dark blue represents the \ha\ broad component, while green is the best-fitting line, which is the combination of all the Gaussian components. Basically, the flux of the \ha\ broad component decreases gradually from the geometric center to the boundary of Region A. At the outskirt, the broad \ha\ component disappears. Approximately 70\% of spaxels within Region A have broad \ha\ components, as marked by a cyan polygon in Figure \ref{fig:morph}(b). Figure \ref{fig:morph}(g) presents \ha+\nii\ spectra for a spaxel at the outskirt of Region A marked by a green cross in Figures \ref{fig:morph}(a) and (b). It is clear the two narrow Gaussian components (blue and red) are enough to describe each observed emission line (black). We suggest one of the two components is dominated by the gaseous rotation disk, while the other one contributes to the abnormal redshifted kinematics. More details about the kinematics of these two components are described in Section \ref{sec:method}. Outside Region A, a single Gaussian component is enough to model each emission line$-$Figure \ref{fig:morph}(h) is an example where the black line is the observed spectrum and the green line is the best-fit single Gaussian model.

\begin{deluxetable}{@{\extracolsep{12pt}}l@{}C@{}C@{}C@{}C@{}C@{}C@{}}[ht!]
    \vspace{-4pt}
    \tabletypesize{\small}
    \centering
    \tablecaption{Coordinates and Projected Separations of Specific Positions.}\label{tab:1}
    \tablehead{\colhead{Position} & \colhead{R.A.(J2000)}                & \colhead{Decl.(J2000)}        & \colhead{$^{*}D_{\rm c}$ ($''$)} & \colhead{$^{\dagger}D_{\rm r}$ ($''$)} & \colhead{$D_{\rm c}$ (kpc)} & \colhead{$D_{\rm r}$ (kpc)}} 
    \startdata
    Galaxy photometric center     & 14$\rm ^h$45$\rm ^m$15.458$\rm ^s$   & +49$\rm ^\circ$26$'$05.21$''$ & $\cdots$                              & 1.02                                   & $\cdots$                         & 0.60  \\
    Geometric center of Region A  & 14$\rm ^h$45$\rm ^m$15.538$\rm ^s$   & +49$\rm ^\circ$26$'$06.74$''$ & 1.72                             & 1.81                                   & 1.02                        & 1.07  \\
    \ha\ broad flux peak          & 14$\rm ^h$45$\rm ^m$15.507$\rm ^s$   & +49$\rm ^\circ$26$'$06.71$''$ & $^{\ddagger}$1.58~               & 1.54                                   & 0.94                        & 0.91  \\
    Radio flux peak               & 14$\rm ^h$45$\rm ^m$15.377$\rm ^s$   & +49$\rm ^\circ$26$'$05.85$''$ & 1.02                             & $\cdots$                                    & 0.60                        & $\cdots$   \\
    \enddata
    \begin{tablenotes}
    \item[] $^{*}$$D_{\rm c}$ represents the distance to the galaxy photometric center.
    \item[] $^{\dagger}$$D_{\rm r}$ represents the distance to the radio flux peak.
    \item[] $^{\ddagger}$We define the positional uncertainty ($\sigma_1$) of the broad \ha\ flux peak as the ratio of the half-light radius to the signal-to-noise ratio within the corresponding radius of the broad \ha\ component. The positional uncertainty ($\sigma_2$) of the galaxy photometric center is given by SDSS. Thus the error of the distance between the broad \ha\ flux peak and the galaxy photometric center is $\sigma=\sqrt{\sigma_1^2+\sigma_2^2}=0.08''$ giving $\sigma_1 = 0.07''$ and $\sigma_2 = 0.04''$. The two positions are obviously different at the $>$3$\sigma$ level.
    \end{tablenotes}
    \vspace{-52pt}
\end{deluxetable} 

\clearpage
\subsection{Multi-Gaussian Fitting Method} \label{sec:method}

We fit each emission line by double Gaussian components using the MPFIT \citep{2009ASPC..411..251M} strategy, which is based on the Levenberg-Marquardt technique to solve the least-squares problem. For each component, \nii$\lambda\lambda$6548,6585, \sii$\lambda\lambda$6718,6732 and \hb\ emission lines are tied to have the same line center as H$\alpha$ in the velocity space, and the line widths are generally allowed to vary in the range of 0.75-1.25 times the \ha\ line width. Since \oiii$\lambda\lambda$4959,5007 does not typically match the profile of the other emission lines \citep{2013ApJ...775..116R}, we fit the \oiii\ lines independently, tying \oiii$\lambda$4959 to have the same line center and line width as \oiii$\lambda$5007 in the velocity space. 

\begin{figure}[ht!]
    \centerline{ \includegraphics[width=1\textwidth]{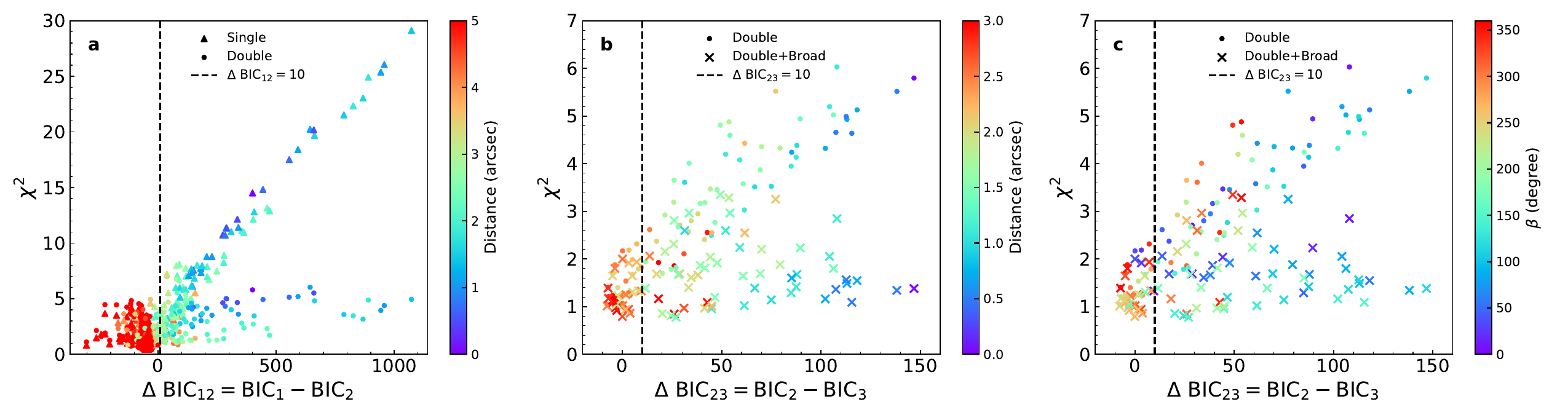} }
    \vspace{0mm}
    \caption{{\bf $\bf \chi^2$ as a function of $\rm\bf \Delta BIC$ for different emission-line models.}
    {\bf (a)} The relation between $\rm \Delta BIC_{12}$ and $\chi^2$ for the \ha+\nii\ wavelength region with $\rm 6529\AA < \lambda < 6610\AA$, where $\rm \Delta BIC_{12}=BIC_1-BIC_2$ is calculated as the difference between the BIC of single ($\rm BIC_1$, triangles) and double ($\rm BIC_2$, dots) Gaussian fittings. 
    {\bf (b)} The same relation but for the double narrow Gaussian and double narrow Gaussian plus a broad \ha\ fitting. $\rm \Delta BIC_{23}=BIC_2-BIC_3$ is measured as the difference between the BIC of emission-line models with ($\rm BIC_3$, crosses) and without ($\rm BIC_2$, dots) broad \ha\ components. 
    In both panels (a) and (b), symbols are color coded by the distance from a certain spaxel to the geometric center of Region A, 
    and the dashed line marks $\rm \Delta BIC=10$.
    {\bf (c)} The same configurations as panel (b), but the symbols are color coded by angle $\beta$ (see Figure \ref{fig:morph}(b) for the definition of $\beta$).}
    \label{fig:bic}
    \vspace{2mm}
\end{figure}

To quantify which spaxels require double Gaussian components, we calculate the Bayesian information criterion \citep[BIC;][]{2022MNRAS.511.5782R} of single ($\rm BIC_1$) and double ($\rm BIC_2$) Gaussian fittings. The difference between single and double Gaussian models, $\rm \Delta BIC_{12}=BIC_1-BIC_2$, is applied to quantify how reasonably the fitting is improved by the double Gaussian model. Figure \ref{fig:bic}(a) shows the relation between $\rm \Delta BIC_{12}$ and the reduced chi-square ($\chi^2$) for the \ha+\nii\ wavelength region with $\rm 6529\AA < \lambda < 6610\AA$, in which triangles (dots) represent spaxels fitted by single (double) Gaussian models. Spaxels with $\rm \Delta BIC_{12}>10$ are selected for double Gaussian fittings \citep{2019MNRAS.487..381S,2021MNRAS.503.5134A}.

Even with the double Gaussian model, there are still many spaxels that have $\chi^2>3$ within Region A due to the existence of broad \ha\ emission. For these spaxels, we refit the \ha+\nii\ emission-line region by further adding a broad \ha\ component and measure its BIC ($\rm BIC_3$). The difference of BIC ($\rm \Delta BIC_{23}=BIC_2-BIC_3$) between double narrow Gaussian fit and double narrow Gaussian plus a broad \ha\ is used to quantify the improvement of the emission-line model by adding a broad \ha\ component. Figure \ref{fig:bic}(b) presents the relation between $\rm \Delta BIC_{23}$ and $\chi^2$ for the \ha+\nii\ wavelength range, where the crosses (dots) represent spaxels fitted with (without) broad \ha\ components. Spaxels with $\rm \Delta BIC_{23}>10$ are selected for refitting with broad \ha\ components. In Figures \ref{fig:bic}(a) and (b), the data points are color coded by the distance from a certain spaxel to the geometric center of Region A. It is clear that the contributions of the broad \ha\ components decrease from the center of Region A to the outskirts. Figure \ref{fig:bic}(c) is the same as Figure \ref{fig:bic}(b), but the only difference is that the data points are color coded by angle $\beta$, which is defined in Figure \ref{fig:morph}(b). We use the red dashed ellipse in the top-right corner of Figure \ref{fig:morph}(b) to represent Region A, where {\it O} is its geometric center. {\it OP}, which is parallel to the galaxy photometric major axis, is set as $\beta = 0^\circ$, with increasing values of $\beta$ along the counterclockwise direction. It is clear from Figure \ref{fig:bic}(c) that spaxels with $\beta < 180^\circ$ (toward the galaxy center) tend to require stronger broad \ha\ components than spaxels with $\beta > 180^\circ$ (toward the outskirts of the galaxy). The dependence of the strength of a broad \ha\ component on $\beta$ could have two explanations: (1) the broad \ha\ component originates from the broad-line region (BLR) of the central BH; (2) the influence of spectral signal-to-noise ratio (SNR) on model fitting. We discuss the first possibility in detail in Section \ref{sec:dust}. Here we focus on the influence of spectral SNR on the detection of the broad \ha\ component.

\begin{figure}[ht!]
    \centerline{ \includegraphics[width=0.8\textwidth]{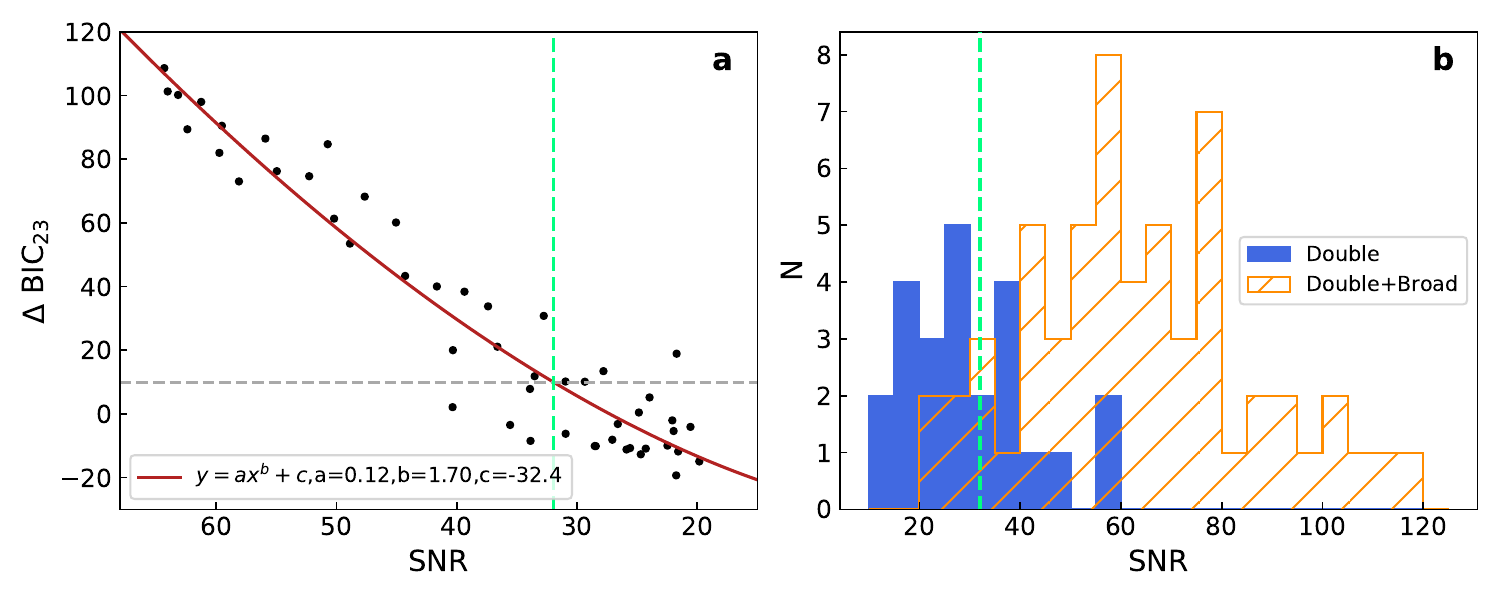} }
    \vspace{0mm}
    \caption{{\bf The effect of SNR on the detection of the broad \ha\ component.}
    {\bf (a)} $\rm \Delta BIC_{23}$ as a function of SNR, where each black dot represents the measurement from one synthetic spectrum, and the red curve is the best-fitted power law of black dots. 
    {\bf (b)} SNR distributions for the observed emission-line spectra within Region A, including spectra modeled by double Gaussian components (the blue shaded area) and double narrow Gaussian plus a broad \ha\ component (the orange hatched area). The green dashed line marks $\rm SNR\sim 32$.}
    \label{fig:snr}
    \vspace{2mm}
\end{figure}

We take the emission-line spectrum around \ha+\nii\ with $\rm 6529\AA < \lambda < 6610\AA$ from the spaxel with maximum broad \ha\ flux, adding Gaussian noise to generate synthetic spectra with different SNRs. We then fit these synthetic spectra using double narrow Gaussian components for each emission line as well as using double Gaussian components plus a broad \ha. Figure \ref{fig:snr}(a) shows $\rm \Delta BIC_{23}$ between these two fitting models as a function of SNR, where each black dot represents the result from one synthetic spectrum. The red line is a power-law fit to the black dots. $\rm \Delta BIC_{23}$ decreases with SNR, and spectra with $\rm SNR\lesssim 32$ (the right side of the green dashed line) have $\rm \Delta BIC_{23}<10$ (below the gray dashed line). We present the SNR distributions for the observed emission-line spectra within Region A in Figure \ref{fig:snr}(b), finding that most spaxels without broad \ha\ components (the blue shaded area) indeed have $\rm SNR\lesssim 32$, while most of spaxels having broad \ha\ components (the orange hatched area) show $\rm SNR\gtrsim 32$. It is clear the SNR of the emission-line spectrum has a strong effect on the detection of the broad \ha\ component, suggesting that the dependence of the strength of broad \ha\ on $\beta$ could be due to the lower SNR for spaxels toward galaxy outskirts ($\beta > 180^\circ$).

\begin{figure}[ht!]
    \centerline{ \includegraphics[width=0.75\textwidth]{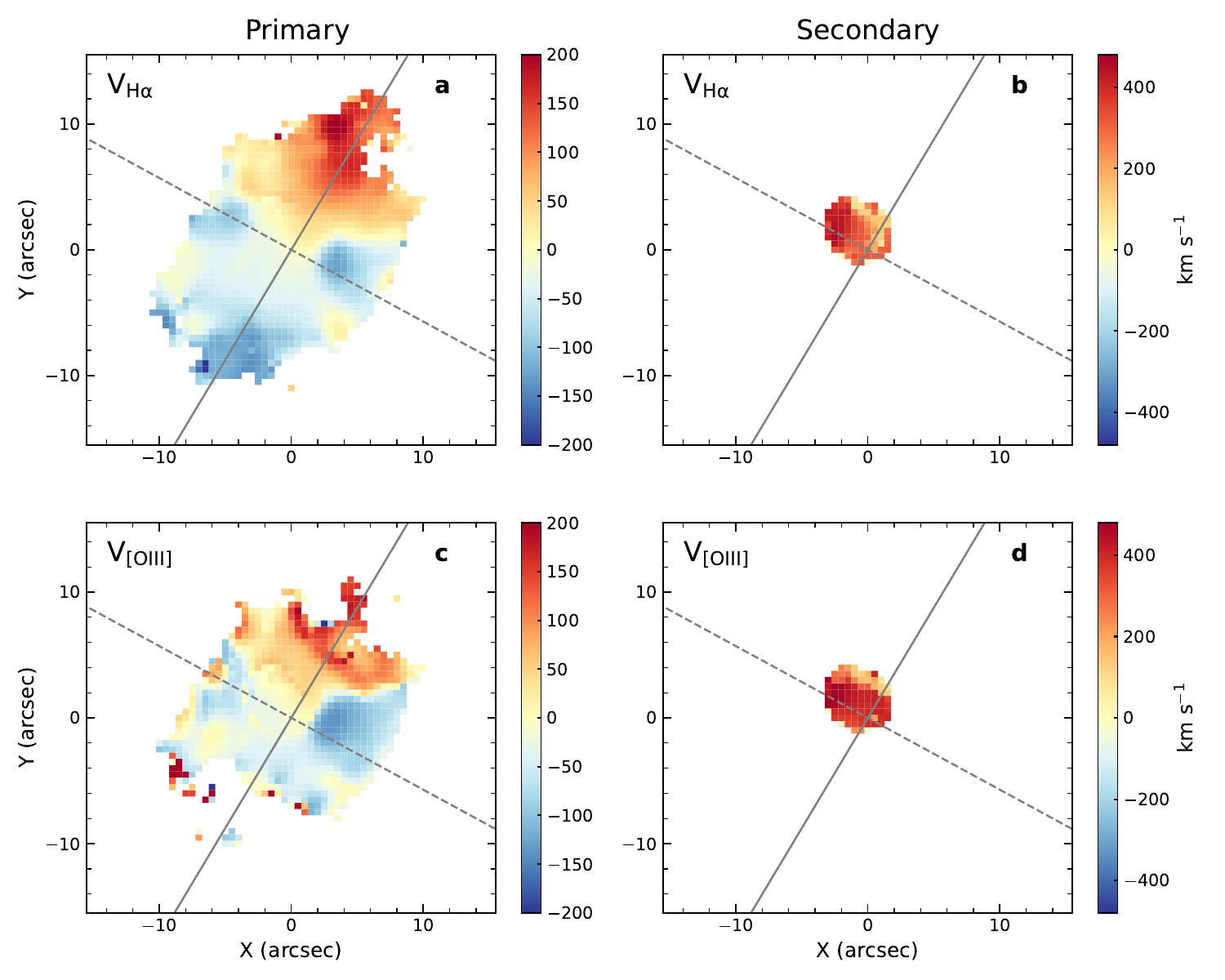} }
    \vspace{0mm}
    \caption{{\bf Decomposed gas kinematics.}
    {\bf (a)} \ha\ velocity field$-$the velocities within Region A are derived from the primary components, while velocities outside Region A are the centroid velocities of single Gaussian models. 
    {\bf (b)} \ha\ velocity field of the secondary components within Region A. {\bf (c),(d)} The same configurations as panels (a),(b) but for \oiii$\lambda$5007.} 
    \label{fig:kine}
    \vspace{2mm}
\end{figure}

Through the emission-line fitting process, we obtain velocity dispersions of $\rm 700-850\ km\ s^{-1}$ for the broad \ha\ components (e.g. dark-blue line in Figure \ref{fig:morph}(d)) which have redshift velocities of $\rm 300-400\ km\ s^{-1}$ relative to the stellar component at the galaxy center. The narrow red components (e.g. red line in Figures \ref{fig:morph}(d)-(g)) show similar redshifted velocities to the broad \ha\ components and velocity dispersions of $\rm 100-200\ km\ s^{-1}$. The narrow blue components (e.g. blue line in Figures \ref{fig:morph}(d)-(g)) have velocity dispersions of $\rm 50-100\ km\ s^{-1}$, maintaining good continuity with the area outside Region A in both velocity and velocity dispersion fields. We define this narrow blue component as a primary one that traces the gaseous rotation disk, and we consider the narrow red Gaussian as a secondary component. Figure \ref{fig:kine} shows the velocity fields for the primary (left) and secondary (right) components. Figure \ref{fig:kine}(a) presents the \ha\ velocity field, where the velocity of the primary component is applied for Region A. Figure \ref{fig:kine}(b) presents the velocity field of the secondary component. Figures \ref{fig:kine}(c) and (d) are similar to Figures \ref{fig:kine}(a) and (b) but for \oiii$\lambda$5007 emission.

\clearpage

\section{Possible origins of the abnormal redshifted gas kinematics} \label{sec:origin}

\subsection{An Off-centered Black Hole?} \label{sec:offcenter}

One natural explanation for the broad \ha\ components with $\rm 700-850\ km\ s^{-1}$ velocity dispersions in Region A is an active BH, while the secondary narrow emission component comes from its narrow-line region (NLR). We calculate the mass of the plausible BH from Region A based on a BH mass ($M_{\rm BH}$) indicator \citep{2015ApJ...801...38W} using the luminosity ($L_{\rm H\alpha}$) and velocity dispersion ($\sigma_{\rm H\alpha}$) of the broad \ha\ component:
\begin{equation}
    M_{\rm BH}=f \times 10^{6.561}{\left(\frac{L_{\rm H\alpha}}{10^{42}\ {\rm erg\ s^{-1}}}\right)}^{0.46}{\left(\frac{\sigma_{\rm H\alpha}}{10^{3}\ {\rm km\ s^{-1}}}\right)}^{2.06}M_{\odot}
\end{equation}
where the virial factor $f=4.47$. $\sigma_{\rm H\alpha}$ and $L_{\rm H\alpha}$ are derived from the stacked spectra of all the spaxels with broad \ha\ components, which are marked by the cyan polygon in Figure \ref{fig:morph}b. The size of the region enclosing spaxels with broad \ha\ components is consistent with $\sim$2.5$''$ $r$-band point spread function (PSF) of SDSS J1445+4926. The mass of the potential off-centered BH is $M_{\rm BH}=(1.3\pm 0.26) \times 10^6M_\odot$ with $\sigma_{\rm H\alpha}= 765\pm 72\ \rm km\ s\rm ^{-1}$ and $L_{\rm H\alpha}= (1.4 \pm 0.03) \times 10^{40}\rm\ erg\ s^{-1}$. $L_{\rm H\alpha}$ is dust attenuation corrected using the Balmer decrement of the primary component based on the assumption of case B recombination \citep{2006agna.book.....O} and the extinction curve from \cite{2000ApJ...533..682C}. The flux peak of the broad \ha\ component is $\sim$0.94 kpc offset from the galaxy center. Coordinates of the galaxy photometric center obtained from SDSS and the \ha\ broad-line flux peak, as well as the projected separation between them, are listed in Table \ref{tab:1}.

\begin{figure}[ht!]
    \centerline{ \includegraphics[width=0.8\textwidth]{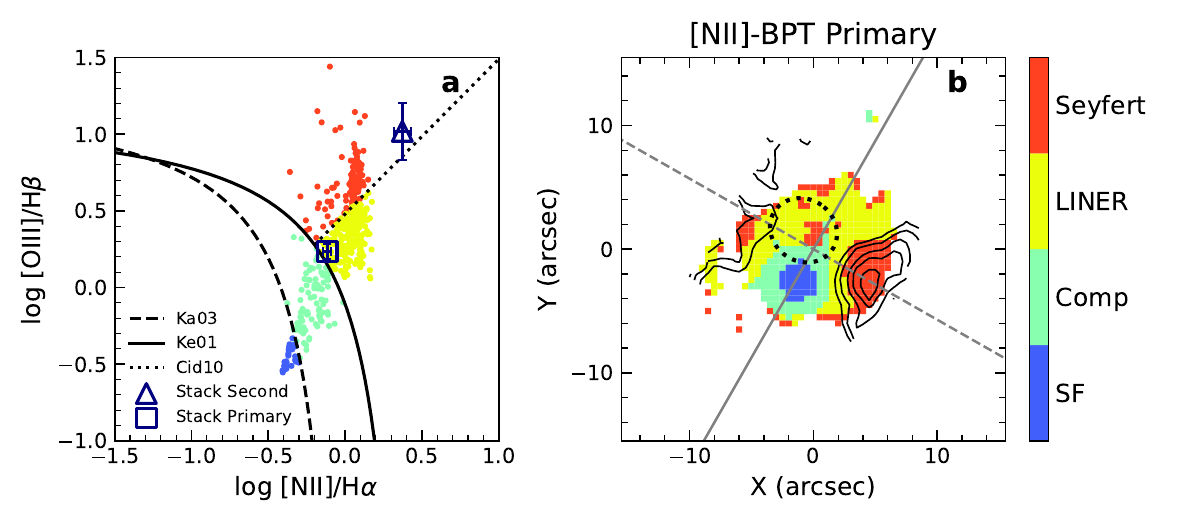} }
    \vspace{0mm}
    \caption{{\bf \nii-BPT diagnostic diagram and spatially resolved BPT map.}
    {\bf (a)} The \nii-BPT diagnostic diagram, where the dashed \citep[Ka03;][]{2003MNRAS.346.1055K} and solid \citep[Ke01;][]{2001ApJ...556..121K} lines are the demarcations for star-forming (SF, blue), composite (Comp, green) and AGNs and the dotted line \citep[Cid10;][]{2010MNRAS.403.1036C} divides AGNs into Seyfert (red) and LINER (yellow). Overlaid square and triangle symbols represent the emission-line ratios of the primary and secondary components from the stacked spectra within Region A. 
    {\bf (b)} The spatially resolved \nii-BPT map for the primary component with the same color code as panel (a). The black contours represent the EW$_{\rm [OIII]}$  with 20\%, 30\%, 50\%, 70\% and 90\% levels, and the photometric major (minor) axis is marked as a gray solid (dashed) line.} 
    \label{fig:bpt}
    \vspace{2mm}
\end{figure}

We apply the traditional \nii\ Baldwin-Phillips-Terlevich (BPT) diagnostic diagram \citep{1981PASP...93....5B} to identify the gas ionization mechanism. Figure \ref{fig:bpt}a shows the \nii/\ha$-$\oiii/\hb\ BPT diagram for spaxels with $\rm SNR>3$ for the four emission lines; the solid \citep{2001ApJ...556..121K} and dashed \citep{2003MNRAS.346.1055K} curves are the demarcations that separate galaxies into AGNs, composite (Comp, green) and star forming (SF, blue). The dotted line in Figure \ref{fig:bpt}(a) is the empirical separation \citep{2010MNRAS.403.1036C} between Seyfert (red) and low-ionization nuclear emission-line region (LINER, yellow). The colored dots are line ratios calculated from the primary components within Region A and single Gaussian fitting outside Region A. We measure the line ratios of the primary and secondary components from the stacked spectra and compare their ionization mechanisms. The square and triangle in Figure \ref{fig:bpt}(a) represent the primary and secondary components, respectively. It is clear that the secondary component that has similar redshift velocity to the broad \ha\ is located at the Seyfert region, while the primary one is located around the boundary of composite and LINERs. Figure \ref{fig:bpt}(b) shows the spatially resolved BPT diagram for the primary component within Region A and single Gaussian fitting outside Region A; the color code is the same as Figure \ref{fig:bpt}(a). We therefore suggest the secondary component comes from the NLR of the off-centered BH. Although gas inflow could also provide redshift kinematics as seen for the secondary component, it is definitely hard to produce the broad \ha\ component with velocity dispersions of $\rm 700-850\ km\ s^{-1}$.

According to the numerical simulations of \cite{2009ApJ...696L..89C,2011ApJ...729...85C} and \cite{2015MNRAS.447.2123C} focusing on the evolution of minor mergers, the secondary BH in the merger remnant will ultimately circularize within the disk of the primary. On the one hand, if the satellite enters the primary galaxy from a coplanar orbit, it will definitely settle into the disk of the primary galaxy; on the other hand, if the satellite enters the primary galaxy from an inclined orbit, the satellite will feel a torque from the primary galaxy, which drags down the inclined orbit to the plane of the primary disk. However, such drag takes a number of orbits to induce significant alignment. Was 49 is a well-studied dual AGN system \citep{1989ApJS...70..271B,1992AJ....104..990M,2017ApJ...836..183S} formed through a minor merger with a stellar mass ratio of the companion (Was 49b) to the primary (Was 49a) between $\sim$1:7 and $\sim$1:15. Was 49b is a bright X-ray source with a broad \ha\ component ($\rm FWHM\sim 6440\pm 60\ km\ s^{-1}$), and it is corotating within the disk of the primary galaxy, suggesting it is beginning to coalesce and circularize in the disk.
In SDSS J1445+4926, the off-centered BH is above the disk plane of the primary galaxy, which appears to be inconsistent with the simulation results. This inconsistency could suggest either SDSS J1445+4926 is still in the dragging process or the dynamics in merging systems is much more complex than the prediction of simulations.

\subsection{Complex Gas Kinematics Triggered by Central AGN Activity?} \label{sec:wind}

\begin{figure}[ht!]
    \centerline{ \includegraphics[width=0.4\textwidth]{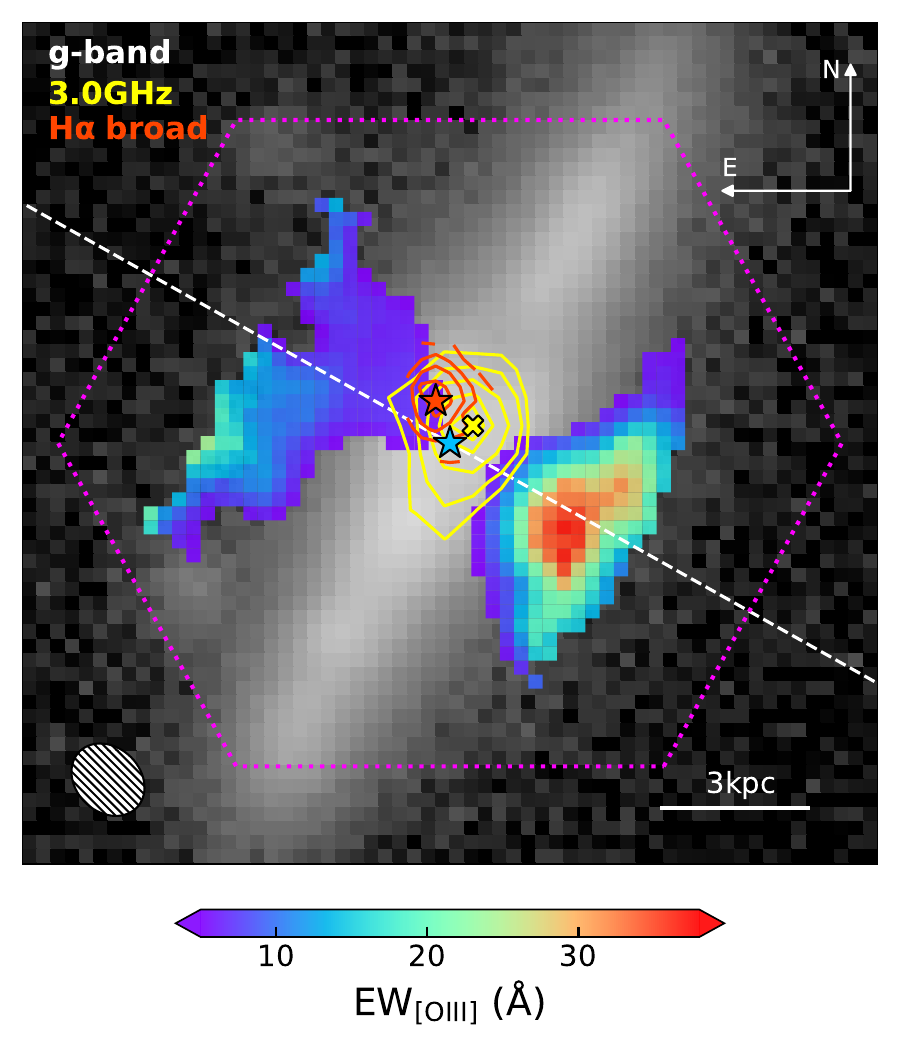} }
    \vspace{0mm}
    \caption{{\bf Distribution of $g$-band flux and \oiii$\lambda$5007\ biconical ionized structure.} 
    The purple hexagon represents the MaNGA bundle.
    The 3.0 GHz VLASS radio continuum is overlaid as yellow contours, and its flux peak is marked as a yellow cross. 
    The VLASS beam size displayed in the lower-left corner is 2.84$''$$\times$2.28$''$, and it has a position angle of 44$^\circ$ (east from north). 
    The flux of the \ha\ broad component within Region A is shown as red contours, and the flux peak is marked as a red star. 
    The overlaid color map shows the EW$_{\rm [OIII]}$.
    The white dashed line is the photometric minor axis, and the blue star marks the photometric center of this galaxy.} 
    \label{fig:radio}
    \vspace{2mm}
\end{figure}

Figure \ref{fig:radio} shows the $g$-band DESI image, overlaid yellow contours are the $\rm 3.0\ GHz$ radio flux distribution from Very Large Array Sky Survey \citep[VLASS;][]{2020PASP..132c5001L}, and red contours represent the flux distribution of broad \ha\ components. The synthesized beam (size: 2.84$''$$\times$2.28$''$; position angle: 44$^\circ$ east from north) of the radio continuum is marked as an ellipse at the bottom-left corner. The position of flux peak in the radio continuum (yellow cross) is $\sim$0.6 kpc from the galaxy center (blue star), while it is $\sim$0.91 kpc away from the flux peak of broad \ha\ components (red star). We list the coordinates of the galaxy photometric center, radio, and broad \ha\ flux peaks, as well as the distance between them in Table \ref{tab:1}. In addition, the flux peak of FIRST $\rm 1.4\ GHz$ emission is close to the galaxy center with a separation of $\sim$155 pc ($\sim$0.26$''$). Although both BH and star-forming activities can contribute to the radio emissions, SDSS J1445+4926 is classified as a radio AGN since its radio luminosity is $>$3$\sigma$ higher than the prediction of the SFR$-$radio luminosity relation for star-forming galaxies \citep{2024arXiv240901279J}, where $\sigma \sim 0.24$ is the intrinsic scatter of this correlation. Considering (1) the position of radio flux peak (including 1.4 and $\rm 3.0\ GHz$) is closer to the galaxy center rather than the broad \ha\ flux peak and (2) the fact that the radio is dominated by the AGN emission, we suggest that there is a central active BH that contributes to the radio emissions. We estimate a central BH mass of $M_{\rm BH}=(9.4\pm 2.7) \times 10^6M_\odot$ based on the $M_{\rm BH}-\sigma_{\star}$ relation \citep{2013ARA&A..51..511K}, where $\sigma_{\star}\sim 90\pm 6\rm\ km\ s^{-1}$ is the average value of stellar velocity dispersions within {\it R}$_e$.

It is noticeable the biconical structure with enhanced EW$_{\rm [OIII]}$ has blueshifted kinematics (regions overlaid with black contours in Figure \ref{fig:morph}(b)) and a Seyfert-like ionization mechanism (regions overlaid with black contours in Figure \ref{fig:bpt}(b)). The maximum velocity dispersion of the \ha\ and \oiii$\lambda$5007 emission lines in the outflow regions is $\rm <130\ km\ s^{-1}$, which is lower than the typical shock-driven velocity dispersions of 150$-$500 km$\rm \ s^{-1}$ \citep{2019ARA&A..57..511K}. We suggest the biconical \oiii$\lambda$5007 ionized structure originates from a central AGN-driven galactic-scale outflow. 

We further estimate a radio spectral index of $\alpha\sim -0.96$ from 3.0 and $\rm 1.4\ GHz$ flux densities; this steep spectrum \citep[$\alpha \lesssim -0.8$;][]{2015ApJ...813..103M} implies the occurrence of a radio jet. \cite{2011ApJ...727...71F} showed that the double-peaked emission lines with $\sim$800 $\rm km\ s^{-1}$ velocity offset seen in ground-based spectra of $\rm Mrk\ 78$ are due to an asymmetric distribution of outflowing gas. 
\cite{2017ApJ...847...41C} and \cite{2016ApJ...830...50M} suggested that velocity offset of $\gtrsim$300 $\rm km\ s^{-1}$ from narrow emission lines could be explained by complex gas kinematics, which is related to the central AGN activity with shocks. The abnormal kinematics in SDSS J1445+4926 are more complex, with the existence of both outflow and jet, which can push away the ISM in the host galaxy creating complex emission-line spectra \citep{2015ApJ...799...72F,2015ApJ...815L...6F,2015ApJ...813..103M}. Although the redshifted secondary component could be driven by the interaction between the jet and ISM, it is difficult to explain the existence of broad \ha\ components in this picture. Follow-up observations of high-resolution multi-waveband imaging and spectra are required to confirm whether the abnormal redshifted components originate from the outflow/jet.

\vspace{20pt}
\subsection{Effect of Dust Extinction?} \label{sec:dust}

\begin{figure}[ht!]
    \centerline{ \includegraphics[width=1\textwidth]{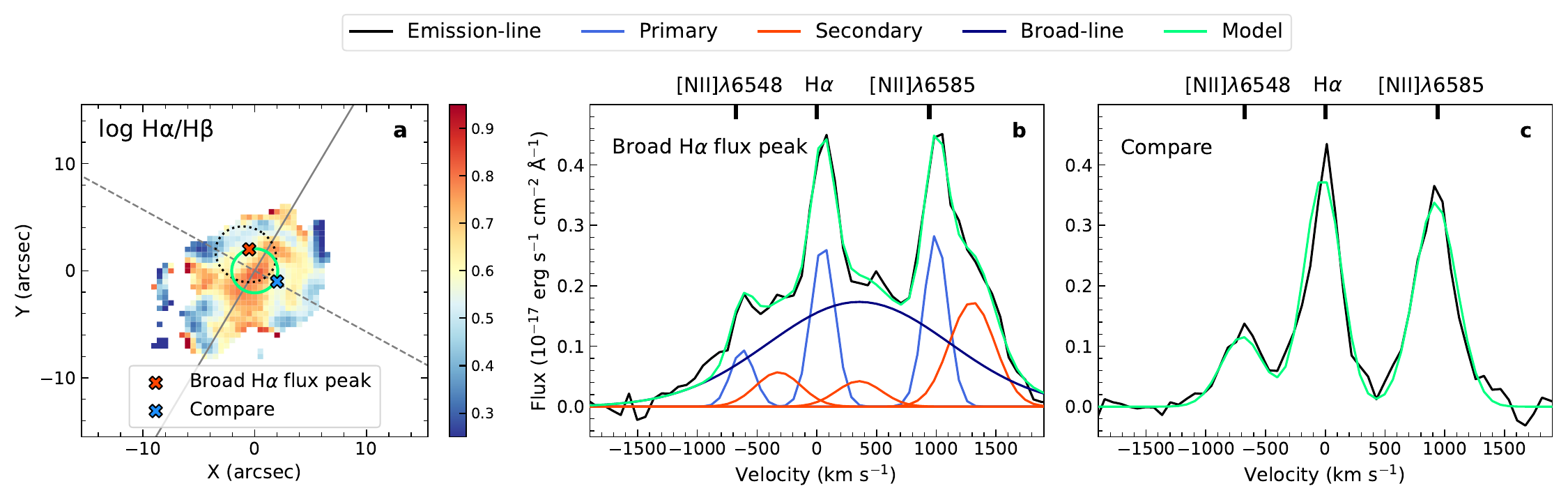} }
    \vspace{0mm}
    \caption{{\bf Comparison of spectra inside and outside Region A.}
    {\bf (a)} The spatially resolved \ha\ to \hb\ flux ratios, where the black dotted ellipse represents Region A and the gray solid (dashed) line is the galaxy photometric major (minor) axis. The green circle is centered on the galaxy photometric center, and its radius is set as the distance from the broad \ha\ flux peak (red cross) to the galaxy center. The spaxel marked with a blue cross outside Region A is located at the green circle and has similar \ha/\hb\ to that of the broad \ha\ flux peak.
    {\bf (b)} The \ha+\nii\ spectrum for the spaxel with maximum broad \ha\ flux (red cross in panel a). The observed emission-line spectrum (black) is well modeled by two narrow Gaussian components (blue and red) plus a broad \ha\ component (dark blue), and the combination of these Gaussian components is the best-fitted model (green).
    {\bf (c)} The \ha+\nii\ spectrum (black) for the comparison spaxel (blue cross in panel (a)); the best-fitted single Gaussian model is shown in green for each emission line.} 
    \label{fig:dust}
    \vspace{2mm}
\end{figure}

Figure \ref{fig:dust}(a) shows the map of nonparametric \ha\ to \hb\ flux ratio (\ha/\hb) which traces the dust extinction of the galaxy. It illustrates that there is a dust lane bisecting the galaxy, which is likely attenuating the flux of both broad and narrow emission lines. In this section, we analyze the possibility that the broad \ha\ component originates from the BLR of the central BH; the decreasing of the broad \ha\ flux from the center of Region A to the galaxy center could be attributed to the fact that the broad-line emitting source is shielded behind the dust curtains. To investigate this possibility, we conduct an examination demonstrated in the following paragraph.

As presented in Figure \ref{fig:dust}(a), we define a circle (with green color) centered on the photometric center of the galaxy; its radius is set as the distance from the broad \ha\ flux peak (red cross) to the galaxy center.
We select a spaxel (blue cross) from this circle with similar \ha/\hb\ to the position of the broad \ha\ flux peak but outside of Region A (black dotted ellipse), and we then compare the \ha+\nii\ emission-line spectra of these two positions. The similar distances of these two spaxels to the galaxy center avoid the effect that the spectral SNR decreases with increasing radius (see Section \ref{sec:method}). Meanwhile, the similar \ha/\hb\ of the two positions excludes the difference on spectral profile induced by dust extinction. In this case, if the broad \ha\ component originates from the galaxy center, we would expect that the two positions have similar spectral structures. The comparison results are shown in Figures \ref{fig:dust}(b) and (c), in which the \ha+\nii\ spectrum of the position of the broad \ha\ flux peak (Figure \ref{fig:dust}(b)) is well modeled by a double narrow Gaussian plus a broad \ha\ component; while for the comparison position (Figure \ref{fig:dust}(c)), a single Gaussian component is enough to model each emission line. Furthermore, the $\rm 300-400\ km\ s^{-1}$ redshifted velocity relative to the stellar component at the galaxy center is difficult to be understood if the broad \ha\ component comes from the BLR of the central BH. Combining these observational evidences, we do not prefer that the broad \ha\ components arise from the BLR of the central BH.

\vspace{60pt}
\section{Conclusions} \label{sec:discuss}

In this work, we investigate an edge-on disk galaxy, SDSS J1445+4926, based on the spatially resolved MaNGA IFU data. This galaxy is notable for its large-scale biconical outflow and the abnormal redshifted gas kinematics located $\sim$1 kpc from its photometric center. Detailed analysis of the emission-line spectra for this region with abnormal redshifted kinematics demonstrates: (i) there exists a broad \ha\ component that has velocity dispersions of about $\rm 700-850\ km\ s^{-1}$ and redshifted velocities of $\rm 300-400\ km\ s^{-1}$ relative to the stellar components at the galaxy center; (ii) double Gaussian components are required to fit each emission line within this abnormal emission region$-$one component follows the kinematics of the rotating gas disk, while the other with similar redshift velocity values to the broad \ha\ component shows Seyfert-like ionization.

One natural explanation of these observational results is a dual active BH candidate with a mass ratio of $\sim$7:1. The central BH with $M_{\rm BH}=(9.4\pm 2.7) \times 10^6M_\odot$ dominates the radio emissions and is driving the galactic-scale biconical outflow. The off-centered BH with $M_{\rm BH}=(1.3\pm 0.26) \times 10^6M_\odot$ is $\sim$1 kpc to the northeast of the galaxy center, which generates the $\rm 300-400\ km\ s^{-1}$ redshifted kinematics through its emissions from the BLR and NLR. The $\rm 300-400\ km\ s^{-1}$ receding velocity of the secondary gas component (Figure \ref{fig:kine}b) definitely rules out the growth of an in situ intermediate-mass BH \citep{1975ApJ...200L.131S,2017IJMPD..2630021M} as the origin of the off-center BH. One natural formation mechanism of this dual AGN candidate is a minor merger. In terms of the morphology of SDSS J1445+4926, the lack of merger remnant features \citep{2021MNRAS.501...14L} in the DESI image could be due to the obscuration from its edge-on disk. In addition, numerical simulations of minor mergers with similar mass ratios demonstrate that there are no observable merger remnant features with surface brightness fainter than 28 $\rm mag\ arcsec^{-2}$ in the $r$-band during the merger processes \citep{2014A&A...566A..97J}.

Aside from the scenario involving a dual AGN candidate discussed above, the complex gas kinematics observed in SDSS J1445+4926 could also originate from an outflow/jet triggered by central AGN activities (e.g. \citealt{2011ApJ...727...71F,2015ApJ...813..103M,2016ApJ...830...50M,2018ApJ...856...93F}) as well as from gas inflows (e.g. \citealt{2017ApJ...847...41C}). However, this picture is much more complicated than the dual AGN system, and we are currently unable to determine the origin of the secondary narrow emission and the broad \ha\ components$-$whether they arise from individual processes (inflow, outflow, or jet) or from the combined contributions of all these mechanisms. Follow-up multi-waveband observations with high spatial or spectral resolutions are necessary to differentiate between these competing scenarios.

\begin{acknowledgments}
    We thank T. Wang, S. Liao, S. Li, and Q. Zhang for their helpful discussions and comments. 
    Y.M.C. acknowledges supports from the National Natural Science Foundation of China (Nos. 12333002 and 11733002), the China Manned Space Project (No. CMS-CSST-2021-A05), and the hospitality of the International Centre of Supernovae (ICESUN), Yunnan Key Laboratory at Yunnan Observatories Chinese Academy of Sciences.
    J.W. acknowledges National Key Research and Development Program of China (No. 2023YFA1607904) and the National Natural Science Foundation of China (Nos. 12033004 and 12221003).
    M.B. acknowledges support from the National Natural Science Foundation of China (No. 12303009). 
    Q.G is supported by  the National Natural Science Foundation of China (Nos. 12192222, 12192220, and 12121003).
    A.M. was supported by the SAO RAS government contract approved by the Ministry of Science and Higher Education of the Russian Federation.
    L.C.H. was supported by the National Science Foundation of China (Nos. 11991052 and 12233001), the National Key Research and Development Program of China (No. 2022YFF0503401), and the China Manned Space Project (Nos. CMS-CSST-2021-A04 and CMS-CSST-2021-A06).
    This work was supported by the research grants from the China Manned Space Project, the second-stage CSST science project ``Investigation of small-scale structures in galaxies and forecasting of observations."

    Funding for the Sloan Digital Sky 
    Survey IV has been provided by the 
    Alfred P. Sloan Foundation, the U.S. 
    Department of Energy Office of 
    Science, and the Participating 
    Institutions. 

    SDSS-IV acknowledges support and 
    resources from the Center for High 
    Performance Computing  at the 
    University of Utah. 
    The SDSS 
    website is www.sdss.org.
    SDSS-IV is managed by the 
    Astrophysical Research Consortium 
    for the Participating Institutions 
    of the SDSS Collaboration including 
    the Brazilian Participation Group, 
    the Carnegie Institution for Science, Carnegie Mellon University, Center for 
    Astrophysics | Harvard \& 
    Smithsonian, the Chilean Participation 
    Group, the French Participation Group, 
    Instituto de Astrof\'isica de 
    Canarias, The Johns Hopkins 
    University, Kavli Institute for the 
    Physics and Mathematics of the 
    Universe (IPMU) / University of 
    Tokyo, the Korean Participation Group, 
    Lawrence Berkeley National Laboratory, 
    Leibniz Institut f\"ur Astrophysik 
    Potsdam (AIP),  Max-Planck-Institut 
    f\"ur Astronomie (MPIA Heidelberg), 
    Max-Planck-Institut f\"ur 
    Astrophysik (MPA Garching), 
    Max-Planck-Institut f\"ur 
    Extraterrestrische Physik (MPE), 
    National Astronomical Observatories of 
    China, New Mexico State University, 
    New York University, University of 
    Notre Dame, Observat\'ario 
    Nacional / MCTI, The Ohio State 
    University, Pennsylvania State 
    University, Shanghai 
    Astronomical Observatory, United 
    Kingdom Participation Group, 
    Universidad Nacional Aut\'onoma 
    de M\'exico, University of Arizona, 
    University of Colorado Boulder, 
    University of Oxford, University of 
    Portsmouth, University of Utah, 
    University of Virginia, University 
    of Washington, University of 
    Wisconsin, Vanderbilt University, 
    and Yale University.
\end{acknowledgments}

\end{document}